 \documentclass[cits]{PoS-pdf}
\usepackage{epsfig}

\usepackage{color}
\let\normalcolor\relax 

\title{
{\small {{DESY 11--253
\\
BI-TP 2011/52}}}\\[2.5cm]
New results for algebraic tensor reduction of Feynman integrals 
}

\ShortTitle{Tensor reduction of Feynman integrals}

\author{Jochem Fleischer\\
Fakult\"at f\"ur Physik, Universit\"at Bielefeld, Universit\"atsstr. 25,  33615
Bielefeld, Germany }
\author{\speaker{Tord Riemann}\\
Deutsches Elektronen-Synchrotron DESY, Platanenallee 6, 15738 Zeuthen, Germany
\\
 E-mail: \email{Tord.Riemann@desy.de}}
\author{Valery Yundin\\Niels Bohr International Academy and Discovery Center, Niels Bohr
Institute, University of Copenhagen, Blegdamsvej 17, DK-2100, Copenhagen,
Denmark}

\abstract{%
We report on some recent developments in  algebraic tensor reduction of one-loop Feynman
integrals.
For 5-point functions, an efficient tensor reduction was worked out recently and is now available as
numerical C++ package, PJFry, covering tensor ranks until five. 
It is free of inverse 5-point Gram
determinants and { inverse small 4-point Gram  determinants are treated by  expansions in
higher-dimensional 3-point functions}.
{By exploiting} sums over signed minors, weighted with scalar products of chords (or, equivalently,
external
momenta), extremely efficient expressions for tensor integrals contracted with
external momenta {were derived}.
The evaluation of 7-point functions is discussed. {In the present approach} one needs  for the reductions
{a $(d+2)$-dimensional scalar 5-point function
} in addition to the usual scalar basis of 1- to 4-point functions in the generic dimension $d=4-2
\epsilon$. 
 When {exploiting} the four-dimensionality of the kinematics, {this} basis is sufficient.
{We indicate how the $(d+2)$-dimensional 5-point function can be evaluated}.
}

\FullConference{ 10th International Symposium on Radiative Corrections (Applications of Quantum Field Theory
to Phenomenology) - Radcor2011\\
September 26-30, 2011\\
Mamallapuram, India}

\begin{document}


\newcommand{\ssL}{{\scriptscriptstyle{L}}}
\newcommand{\ssN}{{\scriptscriptstyle{N}}}
\newcommand{\ssO}{{\scriptscriptstyle{O}}}
\newcommand{\ssQ}{{\scriptscriptstyle{Q}}}
\newcommand{\ssE}{{\scriptscriptstyle{E}}}
\newcommand{\ssD}{{\scriptscriptstyle{D}}}
\newcommand{\ssB}{{\scriptscriptstyle{B}}}
\newcommand{\ssY}{{\scriptscriptstyle{Y}}}

\newcommand{\be}{\begin{equation}}
\newcommand{\ee}{\end{equation}}
\newcommand{\bea}{\begin{eqnarray}}
\newcommand{\eea}{\end{eqnarray}}
\newcommand{\non}{\nonumber}
\newcommand{\nl}{\nonumber \\}

\def \litwo {{\rm{Li_2}}}

\section{Introduction}
In recent years, we worked out a tensor reduction formalism for one-loop Feynman integrals with more than
four legs.
Tensor integrals are
\bea
\label{definition}
 I_{n}^{\mu_1\cdots\mu_R} &=&~\int \frac{d^dk}{i {\pi}^{d/2}}~~\frac{\prod_{r=1}^{R}
k^{\mu_r}}{\prod_{j=1}^{n}c_j},
\eea
with denominators $c_j = (k-q_j)^2-m_j^2 +i \epsilon$ and chords
$q_j$.

An efficient and stable numerical implementation is of high relevance for the description of multi-leg
final states at the LHC.
The state of the art has been regularly summarized at the ``Les Houches Workshops on Physics at TeV
Colliders", see e.g.  \cite{Binoth:2010ra} or the forthcoming proceedings of the 2012 edition.
For a broader context, see also \cite{Blumlein:2010zz,:2008zzy,ZuerichTech.Hochsch.:2010zz}
and references therein.

\begin{figure}[t]
\begin{center}
\includegraphics[width=0.7\textwidth]{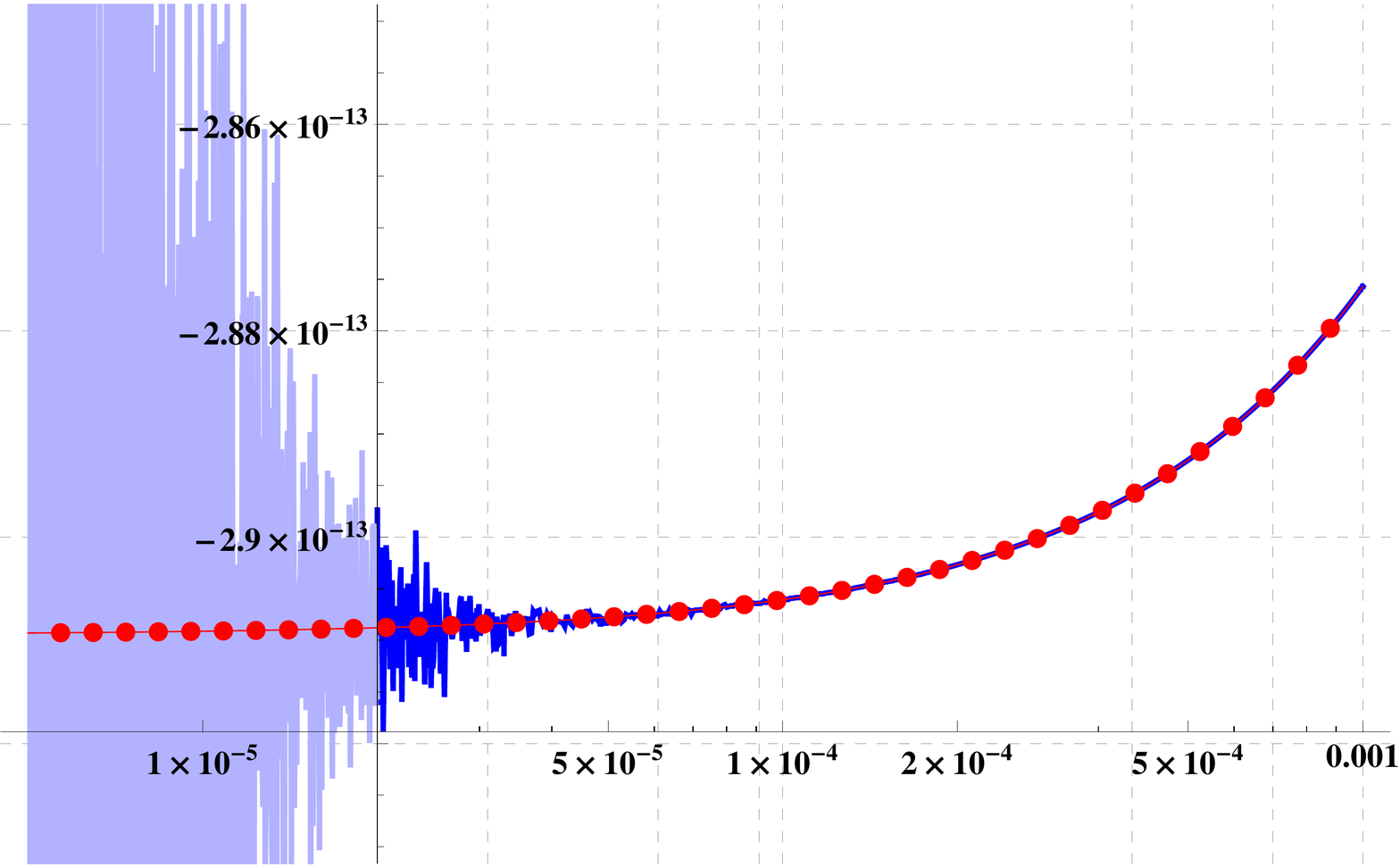}
\caption{The  5-point tensor coefficient  $E_{3333}$ in  the region of a vanishing sub-Gram
determinant.
Blue curve: conventional Passarino-Veltman reduction \cite{Passarino:1978jh}, red curve:
PJFry \cite{Fleischer:2010sq,pjfry:2011-no-url,Fleischer:2011xx?} . Figure copyright: V. Yundin, 2011.
\label{pjfry_fig1}}
\end{center}
\end{figure}

The basics of {our} algebraic approach were formulated in
\cite{Davydychev:1991va,Tarasov:1996br,Fleischer:1999hq}.
In a series of papers, we performed the explicit reduction of 5- and 6-point tensors 
\cite{Fleischer:2007ff,Diakonidis:2008dt,Diakonidis:2008ij,Gluza:2009mj,%
Diakonidis:2009fx,Diakonidis:2010rs,Fleischer:2010sq,Fleischer:2011xy}.
Arbitrary internal masses and external virtualities are allowed. Numerical singularities arising from
inverse  small 5-point Gram determinants are avoided and those from  inverse small 4-point Gram determinants
are
safely evaluated by series of 3-point functions in higher dimensions with improvements by  Pad\'{e}
approximants~\cite{Fleischer:2010sq}.
Based on that, the numerical C++ package PJFry was made available open-source
\cite{Fleischer:2010sq,pjfry:2011-no-url,Fleischer:2011xx?}. 
{An instructive example of typical output for a 6-point function is figure \ref{pjfry_fig1}; for details
on the kinematics see \cite{Fleischer:2011xx?}.}

Since then, the interesting case of contractions of tensor integrals with external momenta was studied.
After such contractions, the resulting scalar quantities are compact linear combinations of the basic scalar
integrals with factorizing, simple combinations of signed minors and scalar products of external momenta
\cite{Fleischer:2011nt,Fleischer:2011bi}.

These developments were summarized at the conference, but because they were described already in earlier
write-ups, we restrict ourselves here to the above quotations.
Quite recently, we also studied the question of how to extend the tensor reductions to more than $n=6$ external
legs.
We again {relied} on the earlier treatment given in \cite{Fleischer:1999hq}.
The reductions may be performed consistently in terms of 1- to 4-point scalar functions, as
was done for $n \leq 6$,  but one has  {in this framework} as
an additional element of the basis the 5-point {function} $I_5^{d+}$ in $d=6-2\epsilon$
dimensions.

This is described in section \ref{seq-7}. 
The additional integral  $I_5^{[d+]}$ is finite, but it is not contained in packages with scalar integrals like 
LoopTools/FF \cite{Hahn:1998yk3,vanOldenborgh:1990yc}, QCDLoop/FF \cite{Ellis:2007qk,vanOldenborgh:1990yc}, or
OneLOop~\cite{vanHameren:2010cp}.
In section \ref{sec-I5d+}, we indicate how to  evaluate this integral  directly.

After the conference we studied {the approach due to \cite{Bern:1992em,Binoth:2005ff}}. It allows
to avoid the use of  $I_5^{[d+]}$. This, together with the effects of contractions with external momenta,  
is studied for 7- and 8-point
tensor integrals in \cite{Fleischer:2011hcmod}, but the details for higher tensor ranks have to be worked out
yet.


\section{Tensor reduction for 7-point functions with recurrence relations\label{seq-7}}
Let us shortly repeat the reasoning for 6-point functions, and then switch to $n >6$.
In   { \cite{Diakonidis:2009fx} }
we have set up the tensor reduction for the { $5$-point} functions of rank $R$, 
expressing any $(5,R)$ pentagon  by a $(5,R-1)$ pentagon plus $(4,R-1)$ boxes:
{
\begin{eqnarray}
\label{tensor5general} %
I_5^{\mu_1  \dots \mu_{R-1} \mu}  &=&I_5^{\mu_1  \dots \mu_{R-1}} Q_0^{\mu} -  \sum_{s=1}^{5}
I_4^{\mu_1  \dots \mu_{R-1},s } Q_s^{\mu} .
\end{eqnarray}
}
In    \cite{Fleischer:2010sq} \footnote{See also \cite{Denner:2005nn}.} we have also given the
corresponding
{ homogeneous} formula for { $6$-point} functions
{
\begin{eqnarray}
\label{tensor6general} %
I_6^{\mu_1  \dots \mu_{R-1} \mu}  &=& -  \sum_{s=1}^{6}
I_5^{\mu_1  \dots \mu_{R-1},s } Q_s^{0,\mu} ,
\end{eqnarray}
}
with $p_6=0$ and
\bea\label{new6}
Q_{s}^{0,\mu}&=&\sum_{i=1}^{5} q_i^{\mu} \frac{ {s0\choose i0}_6} {{0 \choose 0}_6} .
\eea

Now we follow \cite{Fleischer:1999hq}.
For a reduction of $5$-point functions two recursion relations are in principle
sufficient:
\begin{eqnarray}
\label{eq:RR1}
 \nu_j    {\bf j^+} I_n^{{ (d+2)}}
&=&
\frac{1}{{  \left( \right)_n}}
\left[  - {{j \choose 0}_n} +\sum_{k=1}^{n} {j \choose k}_n
   {{\bf k^-} }\right]   {I_n^{{d }}}  
,
\\  
\label{eq:RR2}  
  (d-\sum_{i=1}^{n}\nu_i+1)      I_n^{{ (d+2)}}
  &=&
\frac{1}{ {{\left( \right)_n}}}
  \left[ {{0 \choose 0}_n}
 - \sum_{k=1}^n {0 \choose k}_n {{\bf k^-}} \right]   I_n^{{d}}  
,
\end{eqnarray}
where    
${\bf  i^{\pm}, j^{\pm}, k^{\pm} }$ act by shifting
the indices   {$\nu_i, \nu_j, \nu_k$} by $\pm 1$.
They do not work out for $6$-point functions since {$\left( \right)_n = 0$} for 
{$n \ge 6$}. A further recursion relation, which {does not
decrease dimension}, reads
\begin{eqnarray}
\label{eq:RR3}
 \nu_j    {\bf j^+} I_n^{{ (d)}}
&=&
\frac{1}{{ {0 \choose 0}_n}}\sum_{k=1}^{n} {0j \choose 0k}_n
.
\end{eqnarray}
If reduction of dimension is acceptable, then the above double sum is often reduced by means of
\bea
\sum_{j=1}^{n} \nu_j    {\bf j^+} I_n^{{ (d+2)}} = - I_n^{{
(d)}}.
\label{reddi}
\eea
Further one uses  (4.4) of \cite{Diakonidis:2008ij} together with  (4.5),
\bea
\sum_{i=1}^{5} q_i^{\mu }{0 \choose i}_6=0.
\label{sum6}
\eea
Collecting the above, 6-point functions may be reduced.

\bigskip

For 7-point functions, things are a bit more involed because
we have  ${\left( \right)_7=0}$ and ${{0 \choose j}_7=0}$.
Nevertheless one can apply recursion relations based on
(\ref{eq:RR3}) and (\ref{reddi}), see \cite{Fleischer:1999hq,Diakonidis:2008ij}:

\begin{eqnarray}
\label{eq:Ind1}
{0 \choose 0}_n ~~~I_{n,i}^{[d+]^x}&=&\left[n+1-(d+2x)\right]{0 \choose i}_n  I_{n}^{[d+]^x}+\sum_{r=1}^n {0i
\choose 0r}_n 
I_{n-1}^{[d+]^{(x-1),r}} ,
\\ \nonumber\\
\label{eq:Ind2}
{0 \choose 0}_n {\nu}_{ij}I_{n,ij}^{[d+]^x}&=&\left[n+2-(d+2x)\right]{0 \choose j}_n
I_{n,i}^{[d+]^x}+\left[i\right]^j+
\sum_{r=1, r \ne i}^n {0j \choose 0r}_n  I_{n-1,i}^{[d+]^{(x-1),r}} ,
\\ \nonumber\\
\label{eq:Ind3}
{0 \choose 0}_n {\nu}_{ijk} I_{n,ijk}^{[d+]^x}&=&\left[n+3-(d+2x)\right]{0 \choose k}_n I_{n,ij}^{[d+]^x}+
\left[ij\right]_{red}^k+\sum_{r=1, r \ne ij}^n {0k \choose 0r}_n  I_{n-1,ij}^{[d+]^{(x-1),r}}  ,
\\ \nonumber\\
\label{eq:Ind4}
{0 \choose 0}_n {\nu}_{ijkl} I_{n,ijkl}^{[d+]^x}&=&\left[n+4-(d+2x)\right]{0 \choose l}_n I_{n,ijk}^{[d+]^x}+
\left[ijk\right]_{red}^l+\sum_{r=1, r \ne ijk}^n {0l \choose 0r}_n  I_{n-1,ijk}^{[d+]^{(x-1),r}},
\end{eqnarray}
where
\bea
\left[i\right]^j&=&{0j \choose 0i}_n I_{n}^{[d+]^{(x-1)}},
\\ \nonumber\\
\left[ij\right]^k&=&{0k \choose 0{i}}_n I_{n,j}^{[d+]^{(x-1)}}+
{0k \choose 0{j}}_n I_{n,i}^{[d+]^{(x-1)}}
,
\eea
and the $\left[ij\right]_{{ red}}^k$ is  $\left[ij\right]^k$, but without
repetition of equal indices ${i,j}$.


In  \cite{Fleischer:1999hq} 
an idea of how to proceed for ${7}$-point functions was formulated~-~details,
however, were not given. 
For the {7}-point vector one obtains from \cite{Davydychev:1991va}
\bea
I_7^{\mu}=-\sum_{i=1}^7 q_i^{\mu} I_{7,i}^{[d+]}.
\eea 
The (\ref{eq:Ind3}) with ${ {0 \choose 0}_7=0}$ and ${ {0 \choose
k}_7=0}$ yields
for $i=j=k$ and $x=2$
\bea
{0i \choose 0i}_7 I_{7,i}^{[d+]}+\sum_{r=1,r \ne i}^7 {0i \choose 0r}_7 I_{6,ii}^{[d+],r} = 0
.
\eea
Since for the $6$-point function ${\left( \right)_6 \equiv {r \choose r}_7 =0}$, we have from eq. (55)
of     \cite{Fleischer:1999hq}
\bea
\label{I6ii}
I_{6,ii}^{[d+],r}=\sum_{s=1,s \ne i}^7 \frac{{Rr \choose sr}_7}{{Rr \choose 0r}_7} I_{5,ii}^{[d+],rs}+
 \frac{{Rr \choose ir}_7}{{Rr \choose 0r}_7}  I_{6,i}^{[d+],r},~~~ R=\rm{any ~value~of~ } 0, \ldots, 7. 
\eea
Applying in standard manner the recursion for the $5$-point function
\bea
{\nu}_{ij}I_{5,ij}^{[d+]^l}=-\frac{{0 \choose j}_5}{\left( \right)_5} I_{5,i}^{[d+]^{(l-1)}}+\sum_{s=1}^5
\frac{{s \choose j}_5}{\left( \right)_5} I_{4,i}^{[d+]^{(l-1),s}}+\frac{{i \choose j}_5}{\left( \right)_5}
I_{5}^{[d+]^{(l-1)}}, \nonumber \\
\eea
we replace $I_{5,ii}^{[d+],rs}$ by integrals of the type $I_{5,i}$ and $I_{4,i}^{s}$, the dimension
of which
must not be reduced. Therefore we have to apply recursion (\ref{eq:RR3}) in the form
\bea
{0 \choose 0}_n I_{n,j}=-\left[d-(n+1)\right]{0 \choose j}_n I_n-\sum_{i,k,i \ne k} {0j \choose 0k}_n
I_{n-1,i}^k, ~~~
n=5,4,3,2, \nonumber \\
\eea
i.e. starting at $n=2$ and increasing $n$ step by step, we obtain the desired integrals $I_{5,i}$ and $I_{4,i}^{s}$.
For $n=2$ we have
\bea
{0 \choose 0}_2 I_{2,j}=-\left[d-3)\right]{0 \choose j}_2 I_2-{0j \choose 02}_2 I_{1,1}^2-{0j \choose 01}_2
I_{1,2}^1, \nonumber \\
\eea
with
\bea
I_{1,1}^2&=&\frac{d-2}{2 m_1^2} I_1(m_1^2), 
\\
I_{1,2}^1&=&\frac{d-2}{2 m_2^2} I_1(m_2^2). 
\eea
The problematic case is the integral $I_{6,i}^{[d+],r}$ for which we can write similarly to (\ref{I6ii})
\bea
\label{I6i}
I_{6,i}^{[d+],r}=\sum_{s=1,s \ne i}^7 \frac{{Rr \choose sr}_7}{{Rr \choose 0r}_7} I_{5,i}^{[d+],rs}+
 \frac{{Rr \choose ir}_7}{{Rr \choose 0r}_7} {I_{6}^{[d+],r}} . \nonumber \\
\eea
The ${I_{6}^{[d+],r}}$ cannot so easily be eliminated as in the case of the $6$-point
function, where the vanishing of (\ref{sum6}) was used.
Inserting (\ref{I6i}) into (\ref{I6ii}), there occurs ${Rr \choose ir}_7^2$, i.e. quadratic, so that the
right hand side of  (\ref{sum6}) does not vanish.
\section{Numerical evaluation of higher-dimensional scalar integrals\label{sec-I5d+}} 
Since in the described approach the higher dimensional integral ${I_{6}^{[d+],r}}$ cannot be
eliminated, it is needed to investigate the possibilty for its numerical evaluation --
if  one wants to continue with this approach.
First of all one would reduce the $6$-point function to $5$-point functions by
\bea
I_{6}^{[d+],r}=\sum_{s=1}^7 \frac{{Rr \choose sr}_7}{{Rr \choose 0r}_7} I_5^{[d+],rs},
\eea
such that the problem is shifted to the numerical evaluation of the $I_5^{[d+],rs}$. This, however,
is a well known pathological case, since reducing the $6$-dimensional $5$-point function to $4$-dimensional
$5$- and $4$-point functions, one has
\bea
I_5^{[d+]}=\left[\frac{{0 \choose 0}_5}{\left( \right)_5} I_5-\sum_{s=1}^5\frac{{0 \choose s}_5}{\left( \right)_5} I_4^s \right]
\frac{1}{d-4},
\label{I5d+}
\eea
i.e. for $d=4$ one meets a division by zero, $\frac{0}{0}$. 
In \cite{Jegerlehner:2002es}, however, an interesting approach has been
proposed, which can also be applied to handle
this case. In fact, the idea is to go to even higher dimensions, which provides good
numerical stability.\footnote{Other approaches to the numerical treatment of higher-dimensional scalar
functions are e.g. \cite{Bardin:2000cf,Fleischer:2003rm}.}
%
%
 One-loop $n$-point integrals in arbitrary dimension $d$ can be expressed in standard manner
in terms of Feynman parameters as
\bea
I_n^{(d)}=\Gamma(n-\frac{d}{2}) \int_0^1 dx_1 \dots \int_0^1 dx_{n-1} ~ J_n ~ h_n^{d/2-n},
\eea 
where 
\bea
J_n=x_{n-2}x_{n-3}^2 \cdots x_1^{n-2} ,
\eea
and $h_n$ is a polynomial in the integration variables as well, containing also masses and momenta squared. 
The idea
is to transform these integrals into integrals of higher dimension $D=d-2\epsilon +2 n -2$. 
For
small $\epsilon$ the expansion
of $I_n^{(d+2 n -2)}$  in $\epsilon$ then reads \cite{Jegerlehner:2002es}
\bea
I_n^{(d+2 n -2)}=\Gamma(1+\epsilon) \left[-\frac{s_n}{\epsilon}-s_n-R_n+ O(\epsilon) \right],
\eea
where $s_n$ can be written as
\bea
s_n=\frac{1}{(n+1)!} \sum_{i,j=1}^n Y_{ij} ,
\eea
with 
$Y_{ij}=-(q_i-q_j)^2+m_i^2+m_j^2$
and it is 
\bea
R_n=\int_0^1 dx_1 \dots \int_0^1 dx_{n-1} ~ J_n  ~  h_n~  \ln (h_n).
\eea
Obviously in such an integral no infinities occur anymore in the integrand and numerical integration is
straightforward. It might even be possible and useful to evaluate it analytically.
In our case of a $5$-point function we have $n=5$, i.e. this formula applies for the integral $I_5^{[d+]^4}$.
Applying recursion relations, we can express (\ref{I5d+}) as
\bea
I_5^{[d+]}&=&(d-2)d(d+2) \frac{\left( \right)_5^3}{{0 \choose 0}_5^3} I_5^{[d+]^4} + 
(d-2)d \frac{\left( \right)_5^2}{{0 \choose 0}_5^3} \sum_{s=1}^5 {s \choose 0} I_4^{[d+]^3,s}
\nonumber \\
&&+~
(d-2) \frac{\left( \right)_5}{{0 \choose 0}_5^2} \sum_{s=1}^5 {s \choose 0} I_4^{[d+]^2,s} 
+\frac{1}{{0 \choose 0}_5} \sum_{s=1}^5 {s \choose 0} I_4^{[d+],s}.
\eea
Here we have now the desired representation of $I_5^{[d+]^4}$ in terms of higher-dimensional 
4-point
functions, for which standard recursions can be used to reduce them to integrals in generic
dimension $d$. In this way $7$-point functions could finally be dealt with.

%
%
%

\acknowledgments{%
The authors are grateful to G. Heinrich {and S. Dittmaier} for communication.
TR would like to thank the Organizers of Radcor 2011, D. Indumathi, Prakash
Mathews, Andreas Nyffeler, and V. Ravindran for their warm hospitality and a perfectly organized  conference.
This work is  supported in part by Sonderforschungsbereich/Trans\-re\-gio SFB/TRR 9 of DFG
"Com\-pu\-ter\-ge\-st\"utz\-te Theoretische Teil\-chen\-phy\-sik" and by
European Initial Training Network LHCPHENOnet PITN-GA-2010-264564. 
}

\providecommand{\href}[2]{#2}

\providecommand{\href}[2]{#2}\begingroup\raggedright\endgroup

\end{document}